\documentclass{article}
\usepackage{iclr2027_conference,times}
\iclrfinalcopy

\usepackage{amsmath,amsfonts,bm}

\def\eqref#1{equation~\ref{#1}}

\def\1{\bm{1}}

\DeclareMathAlphabet{\mathsfit}{\encodingdefault}{\sfdefault}{m}{sl}
\SetMathAlphabet{\mathsfit}{bold}{\encodingdefault}{\sfdefault}{bx}{n}

\usepackage{microtype}
\usepackage{booktabs}
\usepackage{array}
\usepackage{tabularx}
\usepackage{graphicx}
\usepackage{float}
\usepackage{placeins}
\usepackage[table]{xcolor}
\usepackage{tikz}
\usetikzlibrary{arrows.meta,positioning}
\usepackage[most]{tcolorbox}
\usepackage{multirow}
\newenvironment{summary}{\begin{tcolorbox}[colback=blue!3,colframe=blue!15,boxrule=0.4pt,arc=1pt,left=4pt,right=4pt,top=3pt,bottom=3pt]}{\end{tcolorbox}}
\usepackage{xspace}
\usepackage{enumitem}
\usepackage{pifont}
\usepackage{amsmath}
\usepackage{amssymb}
\usepackage{hyperref}
\usepackage{url}
\DeclareUnicodeCharacter{2032}{\ensuremath{\prime}}
\definecolor{lightcyan}{RGB}{10,110,150}
\definecolor{lighturlred}{RGB}{251,49,153}
\hypersetup{
  colorlinks=true,
  linkcolor=lightcyan,
  citecolor=lightcyan,
  filecolor=lightcyan,
  urlcolor=lighturlred
}

\newcommand{\benchmark}{{\textsc{SWEPatchTest}}\xspace}

\newcommand{\parabf}[1]{\smallskip\noindent\textbf{#1}\xspace}
\title{SWEPatchTest: Evaluating Software Tests\\Beyond Reproduction}

\author{%
  \textbf{Hua Lu}$^{1}$\thanks{These authors contributed equally to this work.}\quad
  \textbf{Pengcheng Wang}$^{1}$\footnotemark[1]\quad
  \textbf{Han Li}$^{2}$\footnotemark[1]\quad
  \textbf{Zichen Tian}$^{3}$\footnotemark[1]\\[0.5ex]
  \textbf{Yanyu Chen}$^{3}$\quad
  \textbf{Huang He}$^{1}$\thanks{Corresponding authors: Huang He and Bingjin Chen.}\quad
  \textbf{Bingjin Chen}$^{1}$\footnotemark[2]\\[1ex]
  $^1$Baidu Inc. \quad
  $^2$Nanjing University \quad
  $^3$The Chinese University of Hong Kong
}

\begin{document}

\maketitle

\begin{abstract}

Coding agents use generated tests to decide whether a proposed patch satisfies an issue's requirements.
A suite that fails on the buggy program and passes after applying the reference patch can still accept an incorrect patch or reject a correct patch.
We introduce \benchmark\footnote{\raggedright Our data are available at \url{https://huggingface.co/datasets/anonymous-swepatchtest/SWEPatchTest}.}, a test-generation benchmark with 619 tasks from SWE-bench Pro, SWE-bench Multilingual, and DeepSWE and 82,579 candidate patches collected from coding-agent trajectories.
By replaying each suite on the buggy and patched codebases, \benchmark jointly measures issue reproduction, acceptance of correct patches, and rejection of incorrect patches.
Joint Validation Rate (JVR), which requires all three criteria on the same task, exposes failures hidden by single-reference evaluation: the best-performing configuration achieves only 9.21\% JVR despite reproducing the reference transition on 59.29\% of tasks.
We further use validation scores to select and scale demonstrations for supervised fine-tuning (SFT). The resulting training recipe approximately triples JVR and improves downstream repair across three SWE-bench benchmarks.
Patch discrimination thus provides both a diagnostic measure of test quality and an execution-based signal for curating agent training data.


\end{abstract}

\section{Introduction}
\label{sec:introduction}

Large language model (LLM) coding agents address software issues by exploring repositories, editing code, and using execution feedback to revise their patches~\citep{NEURIPS2024_5a7c9475,xia2024agentlessdemystifyingllmbasedsoftware,wang2025openhandsopenplatformai,li2026contextbenchbenchmarkcontextretrieval}.
SWE-bench and SWE-bench Pro assess the resulting patches with executable test suites~\citep{jimenez2024swebench,deng2025swebenchproaiagents}.
During repair, however, agents generate their own tests to distinguish correct patches (passing official evaluation) from incorrect patches~\citep{pan2026prometheuslonghorizoncodebasenavigation}.

\parabf{Limitations of existing benchmarks.}
Table~\ref{tab:compare_benches} contrasts the scale, language coverage, and evaluation criteria of test-generation (TestGen) benchmarks.
Pool sizes exclude reference patches and include mutants as negative candidates; checks denote unreported sizes, and crosses denote no pool in the main evaluation.
SWT-Bench and TDD-Bench Verified evaluate generated suites through a fail-to-pass (F2P) transition: failure on the buggy program followed by success on the reference patch~\citep{NEURIPS2024_94f093b4,ahmed2024tddbenchverifiedllmsgenerate}.
This establishes reference recognition but leaves two complementary requirements untested: \textbf{(i) accepting correct patches} and \textbf{(ii) rejecting incorrect patches}.
We refer to implementation-dependent rejection of correct patches as \emph{solution bias} and insufficient checking of required behavior as \emph{weak specification}; either can coexist with a successful reference transition.

Test-suite overfitting lets patches satisfy available tests while retaining faults~\citep{Smith_2015}.
Evaluating both candidate classes reveals whether a test captures the issue's requirements beyond one reference patch.
Prior work already uses generated tests for patch filtering and evaluates suites against multiple patches or mutations~\citep{NEURIPS2024_94f093b4,sonwane-etal-2026-omnicode,sun2026swemutationllmsgeneratereliable}.
We ask how reliably these tests accept correct patches and reject incorrect patches drawn from diverse coding-agent trajectories.

\providecommand{\cmark}{\textcolor{green!50!black}{\ding{51}}}
\providecommand{\xmark}{\textcolor{red!80!black}{\ding{55}}}

\begin{table}[htbp]
\centering
\caption{
Comparison of test-generation benchmarks.
}
\label{tab:compare_benches}
\small
\renewcommand{\arraystretch}{1.10}
\setlength{\tabcolsep}{4pt}
\resizebox{\linewidth}{!}{%
\begin{tabular}{lccccrlc}
\toprule
Benchmark & F2P & Correct pool & Incorrect pool & Joint Validation & Tasks & Test scale & Multilingual \\
\midrule
TestEval & \xmark & \xmark & \xmark & \xmark & 210 & Function & \xmark \\
TestGenEval & \xmark & \xmark & \cmark & \xmark & 1,210 & File & \xmark \\
CLOVER & \xmark & \xmark & \xmark & \xmark & 845 & Block/file & \xmark \\
MultiFileTest & \xmark & \xmark & \xmark & \xmark & 60 & Multi-file & 3 \\
SWT-Bench & \cmark & \xmark & \xmark & \xmark & 1,983 & Repo. & \xmark \\
TDD-Bench Verified & \cmark & \xmark & \xmark & \xmark & 449 & Repo. & \xmark \\
OmniCode & \xmark & \xmark & 1,109 & \xmark & 396 & Repo. & 3 \\
SWE-Mutation & \cmark & \xmark & 2,636 & \xmark & 800 & Repo. & 10 \\
\midrule
\textbf{\benchmark} & \cmark & \textbf{27,749} & \textbf{54,830} & \cmark & \textbf{619} & Repo. & \textbf{10} \\
\bottomrule
\end{tabular}%
}
\end{table}

\parabf{\benchmark: from one reference to a patch distribution.}
\benchmark pairs 619 tasks from SWE-bench Pro, SWE-bench Multilingual, and DeepSWE with candidate patches~\citep{deng2025swebenchproaiagents,NEURIPS2025_8b86cf5a,huang2026deepswemeasuringfrontiercoding}.
We select tasks with suitable reference tests and nontrivial reproduction difficulty, then collect and deduplicate patches from diverse model and agent configurations.
The benchmark contains 82,579 candidate patches, split into correct and incorrect pools; Section~\ref{sec:benchmark} reports the class counts.

\begin{figure}[t]
\centering
\includegraphics[width=\linewidth]{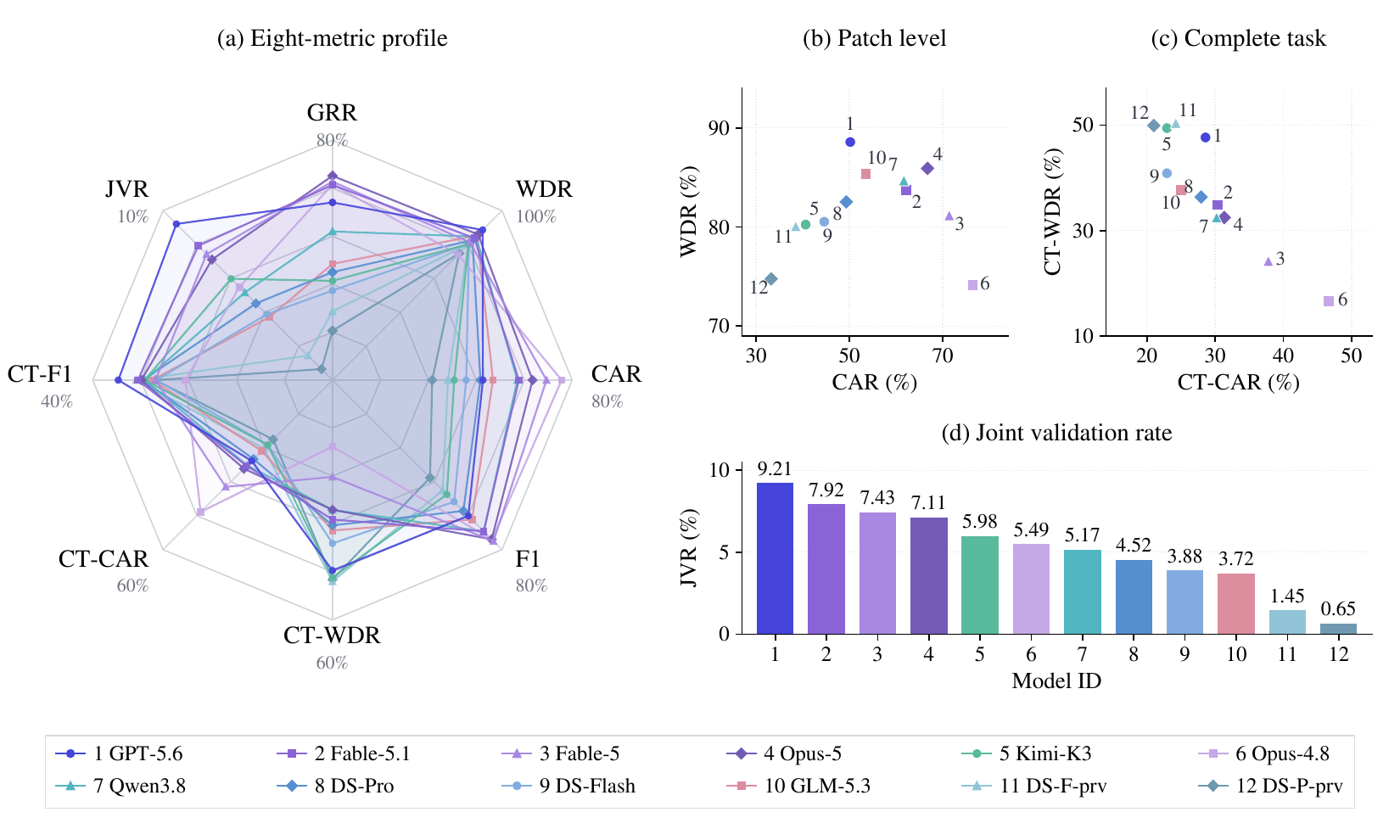}
\caption{
Patch-discrimination profiles of coding-agent configurations on \benchmark.
}
\label{fig:leaderboard_views}
\end{figure}

Each generated suite is replayed on the buggy codebase and codebases obtained by applying the reference or each candidate patch to check \textbf{(i) issue reproduction}, \textbf{(ii) correct-patch acceptance}, and \textbf{(iii) incorrect-patch rejection}.
Gold Recognition Rate (GRR), Correct Patch Acceptance Rate (CAR), and Wrong Patch Detection Rate (WDR) measure these requirements, respectively.
Complete-task rates require success on every candidate in a class; we propose Joint Validation Rate (JVR) to require both complete acceptance and complete rejection together with reference recognition on the same task.
Figure~\ref{fig:leaderboard_views} summarizes these measures and their harmonic means, separating recognition of the reference patch from discrimination over candidate patches.
The radar axes start at zero and scale linearly to the labeled percentage limits; all panels share model IDs and colors, with abbreviations listed in Appendix~\ref{app:agent_configurations}.

\parabf{Takeaways.}
We evaluate 12 coding-agent configurations and investigate whether patch-discrimination scores can guide supervised fine-tuning (SFT), yielding three takeaways.
\ding{182}~\textbf{Joint validation exposes failures hidden by reference recognition.} Even the strongest configuration achieves JVR of only 9.21\%, compared with its 59.29\% GRR, revealing weaknesses that single-reference evaluation leaves unexamined.
\ding{183}~\textbf{Rejecting incorrect patches does not ensure reliable validation.} The configuration with the highest WDR (88.57\%) accepts only 50.20\% of correct patches, revealing why test quality must account for both failure detection and acceptance of correct patches.
\ding{184}~\textbf{Validation-guided test-generation improves both testing and repair.} Selecting and scaling test-generation supervision approximately triples JVR on \benchmark and achieves competitive performance across downstream repair benchmarks.

\parabf{Contributions.}
\begin{enumerate}[label=\textbf{(\roman*)},leftmargin=1.8em,itemsep=2pt,topsep=3pt,parsep=0pt]
\item We construct \benchmark, a repository-level test-generation benchmark with 619 tasks across ten programming languages and 82,579 candidate patches, enabling evaluation over distributions of diverse correct and incorrect patches.

\item We introduce Joint Validation Rate (JVR) and an execution-based protocol that formalize reliable test validation as jointly requiring issue reproduction, correct-patch acceptance, and incorrect-patch rejection on the same task; evaluation of 12 coding-agent configurations exposes limitations obscured by reference recognition alone.

\item We develop a validation-guided test-generation pipeline that selects demonstrations using WDR, CAR, and JVR and show that scaling this supervision improves test-generation quality on \benchmark and downstream repair on SWE-bench Verified~\citep{jimenez2024swebench}, SWE-bench Multilingual~\citep{NEURIPS2025_8b86cf5a}, and SWE-bench Pro~\citep{deng2025swebenchproaiagents}.
\end{enumerate}

\begin{figure}[H]
\centering
\includegraphics[width=\linewidth]{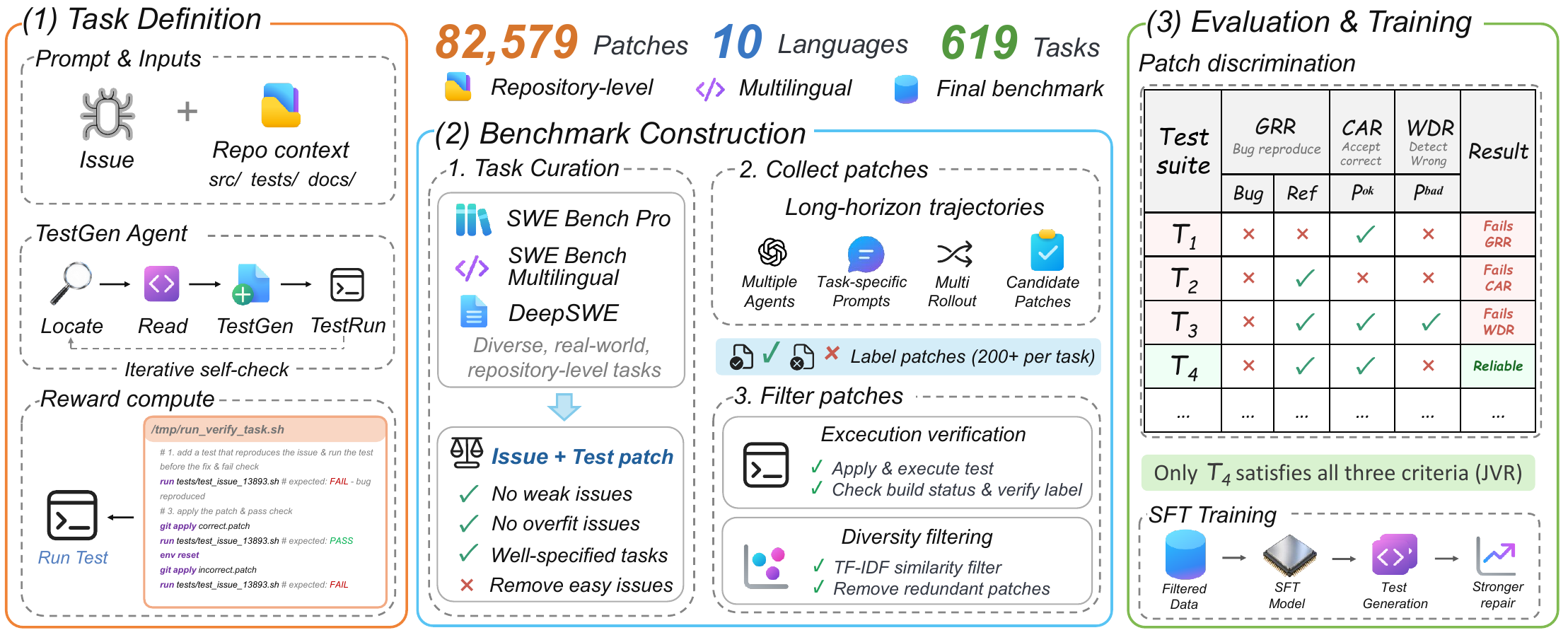}
\caption{
Overview of test generation, benchmark construction, and SFT in \benchmark.
}
\label{fig:pipeline}
\end{figure}

Figure~\ref{fig:pipeline} provides an overview of test generation, benchmark construction, and validation-guided SFT in \benchmark.

\section{\benchmark: A Benchmark for Patch Discrimination}
\label{sec:benchmark}

\benchmark evaluates a generated test suite against multiple patches for the same issue, extending reference reproduction to judgments across implementations.
We construct the benchmark in three stages: \textbf{(1) task curation} establishes reference-test quality and reproduction difficulty; \textbf{(2) candidate patch collection} gathers candidate patches; and \textbf{(3) candidate verification and selection} forms the correct and incorrect pools.
Each task contains an issue, a buggy program, a reference patch, and correct and incorrect patch pools.

\subsection{Step 1: Task Curation}
\label{sec:selection}

We begin with 1,144 repository-level tasks: 731 from SWE-bench Pro, 300 from SWE-bench Multilingual, and 113 from DeepSWE~\citep{deng2025swebenchproaiagents,NEURIPS2025_8b86cf5a,huang2026deepswemeasuringfrontiercoding}.
Task curation targets two properties: \textbf{(i) the reference tests should appropriately enforce the behavior required by the issue}, and \textbf{(ii) reproducing the fail-to-pass transition should remain nontrivial for current test-generation configurations}.
Applying these filters leaves 619 tasks for evaluation; Table~\ref{tab:filter_funnel} summarizes the resulting benchmark composition by source.

\parabf{Reference-test quality.}
We apply an LLM-based audit that jointly examines the issue description, reference patch, and official tests.
The audit excludes tasks with weak or implementation-dependent reference assertions, addressing test-oracle limitations~\citep{Barr_2015,chu2026terminalworldbenchmarkingagentsrealworld}.
We additionally conduct a manual review of the filtering results before finalizing task eligibility.
Appendix~\ref{app:task_curation} provides the judging criteria and further details.

\parabf{Reference-transition difficulty.}
We further filter tasks using a heterogeneous panel of test-generation configurations, removing issues whose fail-to-pass transition is consistently reproduced across the panel.
This step removes tasks that are broadly easy under the panel and yields a challenge set with nontrivial reference-transition difficulty.
Appendix~\ref{app:task_curation} summarizes the task-curation procedure.

\subsection{Step 2: Candidate Patch Collection}

We collect candidate patches from diverse model families and coding-agent harnesses; Appendix~\ref{app:pools} lists the model inventory in Table~\ref{tab:collection_models}.
For each task, we retain patches and their initial execution outcomes from coding-agent trajectories, capturing diverse implementations in both classes for evaluation beyond the reference patch.

\subsection{Step 3: Candidate Verification and Selection}
We apply each candidate patch and rerun the official task evaluation: passing patches form the correct pool and failing patches the incorrect pool.
Within each task and correctness class, we remove duplicate and near-duplicate patches and select diverse subsets to preserve implementation diversity.
The reference patch is retained as a separate evaluation target rather than included in the candidate pool.
Appendix~\ref{app:pools} details similarity filtering, diversity selection, and final curation.

\subsection{\benchmark Statistics}
\label{sec:statistics}

The final benchmark contains \textbf{619 tasks across ten programming languages}: 342 from SWE-bench Pro, 167 from SWE-bench Multilingual, and 110 from DeepSWE.
It contains \textbf{82,579 candidate patches}: 27,749 correct patches and 54,830 incorrect patches, averaging 44.83 and 88.58 per task, respectively (Table~\ref{tab:pool_stages}).
These pools support evaluation of both correct-patch acceptance and incorrect-patch rejection.

\subsection{\benchmark Evaluation Protocol}
\label{sec:protocol}
\label{sec:metrics}

\parabf{Codebases, patches, and test execution.}
Following the notation of SWT-Bench~\citep{NEURIPS2024_94f093b4}, let $R_i$ denote the buggy codebase for task $i$ and $R_i\circ p$ the codebase obtained by applying patch $p$.
Each task provides a reference patch $p_i^{ref}$ and correct and incorrect patch pools $P_i^{ok}$ and $P_i^{bad}$.
For a test $s$ and codebase $R$, $\operatorname{exec}(s,R)\in\{\mathrm{P},\mathrm{F},\bot\}$ records a completed pass, a stable behavioral failure, or an unresolved outcome.
We extend this notation to a submitted nonempty suite $T_i$: $\operatorname{exec}(T_i,R)=\mathrm{P}$ when all tests pass, $\mathrm{F}$ for a completed run with at least one stable behavioral failure, and $\bot$ for missing or unresolved executions, including infrastructure-only errors.
We replay the same suite unchanged on clean copies of $R_i$, $R_i\circ p_i^{ref}$, and each $R_i\circ p$; Appendix~\ref{app:suite} gives the replay protocol.

\parabf{Measuring acceptance and rejection.}
We define reference reproduction $g_i$, candidate acceptance $a_i(p)$, and candidate rejection $r_i(p)$ directly from execution outcomes:
\begin{equation}
\begin{aligned}
g_i&=\mathbf{1}[\operatorname{exec}(T_i,R_i)=\mathrm{F}\ \wedge\ \operatorname{exec}(T_i,R_i\circ p_i^{ref})=\mathrm{P}],\\
a_i(p)&=\mathbf{1}[\operatorname{exec}(T_i,R_i\circ p)=\mathrm{P}],\qquad
r_i(p)=\mathbf{1}[\operatorname{exec}(T_i,R_i\circ p)=\mathrm{F}].
\end{aligned}
\label{eq:execution_indicators}
\end{equation}
An unresolved outcome earns neither acceptance nor rejection credit.
Unresolved executions remain in the score denominators, so failures to produce a usable verdict cannot improve measured discrimination.
We replay inconsistent outcomes and require stable behavioral evidence before crediting rejection (Appendix~\ref{app:suite}).
The fractions of incorrect patches rejected and correct patches accepted are
\begin{equation}
w_i=\frac{\sum_{p\in P_i^{bad}}r_i(p)}{|P_i^{bad}|},
\qquad
c_i=\frac{\sum_{p\in P_i^{ok}}a_i(p)}{|P_i^{ok}|}.
\label{eq:candidate_rates}
\end{equation}
Gold Recognition Rate (GRR), Wrong Patch Detection Rate (WDR), and Correct Patch Acceptance Rate (CAR) average these outcomes over the same $N=619$ tasks:
\begin{equation}
\mathrm{GRR}=\frac{1}{N}\sum_{i=1}^{N}g_i,
\qquad
\mathrm{WDR}=\frac{1}{N}\sum_{i=1}^{N}w_i,
\qquad
\mathrm{CAR}=\frac{1}{N}\sum_{i=1}^{N}c_i.
\label{eq:discrimination}
\end{equation}
All tasks receive equal weight, and the candidate pools are fixed across evaluated configurations.
Appendix~\ref{app:metrics} details the scoring conventions.

\parabf{Complete-task and balanced scores.}
Complete-task (CT) rates count tasks with a correct verdict for every candidate in the corresponding class: CT-WDR requires rejecting all incorrect patches, and CT-CAR requires accepting all correct patches.
F1 is the harmonic mean of WDR and CAR, and CT-F1 is the harmonic mean of CT-WDR and CT-CAR; Appendix~\ref{app:metrics} gives their formulas.

\parabf{Requiring joint validation.}
We propose \textbf{Joint Validation Rate (JVR)} to measure whether a suite jointly satisfies all three validation requirements on the same task: reproducing the issue, rejecting every incorrect patch, and accepting every correct patch.
\begin{equation}
\mathrm{JVR}=\frac{1}{N}\sum_{i=1}^{N}\mathbf{1}[g_i=1\ \wedge\ w_i=1\ \wedge\ c_i=1].
\label{eq:jvr}
\end{equation}
We rank configurations by JVR because it requires all three capabilities on the same task; the other rates diagnose its components.
All comparisons use fixed candidate pools, since larger or more diverse pools make complete validation harder.

\section{How Reliably Do Generated Tests Validate Patches?}
\label{sec:experiments}

To evaluate generated tests as patch validators and examine the training value of their execution outcomes, we investigate the following research questions (RQs):
\begin{itemize}[leftmargin=4mm,itemsep=0.05mm,topsep=2pt,parsep=0pt,partopsep=0pt]
    \item \textbf{RQ1: Patch-Discrimination Performance and Failure Analysis.}
    \textit{How reliably do generated tests reproduce issues, accept correct patches, and reject incorrect patches on the same task, and what failure modes emerge?}
    \item \textbf{RQ2: Selecting Test-generation Demonstrations.}
    \textit{Does selecting and scaling SFT demonstrations using patch-discrimination scores improve test generation beyond reference-based selection?}
    \item \textbf{RQ3: Transfer to Downstream Repair.}
    \textit{Does validation-guided test-generation supervision improve repair success on SWE-bench Verified, Multilingual, and Pro?}
\end{itemize}

\parabf{Shared evaluation setup.}
All \benchmark evaluations use the same 619 tasks, reference patches, candidate pools, and labels (Section~\ref{sec:metrics}).
Figure~\ref{fig:leaderboard_views} compares 12 main configurations, Table~\ref{tab:main_results} reports eight representative configurations, Table~\ref{tab:sft_results} reports four SFT configurations, and other runs support execution audits.
Both tables report percentages; the main table bolds column maxima among the displayed configurations.
We evaluate models with their native coding agents, using OpenHands for the DeepSeek configurations; Appendix~\ref{app:agent_configurations} details the agent setup.
The comparison characterizes complete agent configurations, including their scaffolds and reasoning settings.

\subsection{RQ1: Patch-Discrimination Performance and Failure Analysis}
\label{sec:rq1}


















\begin{table*}[t]
\centering
\caption{
Patch-discrimination performance of representative coding agents on \benchmark.
}
\label{tab:main_results}

\begingroup
\small
\renewcommand{\arraystretch}{1.14}

\begin{tabular*}{\textwidth}{
@{\hspace{0.01\textwidth}}
l
@{\extracolsep{\fill}}
cccccccc
@{\hspace{0.02\textwidth}}
}
\toprule

\multirow{2}{*}{Model}
& \multicolumn{1}{c}{Reference}
& \multicolumn{3}{c}{Patch-level}
& \multicolumn{4}{c}{Complete-task} \\[-2pt]

\cmidrule(lr){2-2}
\cmidrule(lr){3-5}
\cmidrule(lr){6-9}

&
GRR
& WDR
& CAR
& F1
& CT-WDR
& CT-CAR
& CT-F1
& JVR \\[-1pt]

\midrule

GPT-5.6-sol
& 59.29
& \textbf{88.57}
& 50.20
& 64.08
& 47.66
& 28.59
& \textbf{35.74}
& \textbf{9.21} \\

Claude-Fable-5.1
& 65.11
& 83.74
& 62.19
& 71.37
& 34.89
& 30.37
& 32.47
& 7.92 \\

Claude-Opus-5
& \textbf{68.17}
& 85.91
& \textbf{66.76}
& \textbf{75.13}
& 32.47
& \textbf{31.34}
& 31.89
& 7.11 \\

Kimi-K3
& 33.12
& 80.24
& 40.64
& 53.95
& \textbf{49.43}
& 22.94
& 31.34
& 5.98 \\

Qwen3.8-max
& 49.60
& 84.69
& 61.68
& 71.38
& 32.47
& 30.21
& 31.30
& 5.17 \\

DeepSeek-V4-Pro
& 36.03
& 82.53
& 49.36
& 61.77
& 36.35
& 27.95
& 31.60
& 4.52 \\

DeepSeek-V4-Flash
& 29.89
& 80.53
& 44.62
& 57.42
& 40.87
& 22.94
& 29.39
& 3.88 \\

GLM-5.3
& 38.77
& 85.36
& 53.55
& 65.81
& 37.64
& 25.04
& 30.07
& 3.72 \\

\bottomrule
\end{tabular*}

\endgroup
\end{table*}

\parabf{Reliable patch discrimination remains far from solved.}
Strong performance on the reference patch leaves a substantial gap in validation across candidate patches (Table~\ref{tab:main_results}).
GPT-5.6-sol achieves the highest JVR, 9.21\%, despite recognizing the reference transition on 59.29\% of tasks.
Claude Opus 5 has the highest GRR at 68.17\%, but its JVR is 7.11\%, below GPT-5.6-sol.
The change in ranking shows that strong reference recognition does not necessarily translate into reliable discrimination among candidate patches.

\parabf{Uncertainty in joint-validation performance.}
A paired, source-stratified bootstrap~\citep{Efron_1979} with 10,000 replicates yields a 95\% confidence interval of $[6.95,11.47]\%$ for GPT-5.6-sol's JVR (Appendix~\ref{app:statistical_reliability_rq1}).
Each resample preserves the three source sizes and uses identical task indices across configurations, retaining the source composition and task pairing for comparisons.
Its observed lead over Claude-Fable-5.1 and Claude-Opus-5 is not statistically significant under exact two-sided McNemar tests with Holm correction ($p=.350$ and $.222$, respectively).
Thus, the evidence establishes low joint-validation performance more clearly than a definitive ordering among the leading configurations.

\parabf{Detecting failures and preserving correct patches are distinct requirements.}
Candidate-level scores reveal why reference success is insufficient: a suite can miss incorrect patches, reject correct patches, or do both on different tasks.
GPT-5.6-sol has the highest WDR at 88.57\%, yet accepts only 50.20\% of correct patches on average.
Among the 12 configurations in Figure~\ref{fig:leaderboard_views}, Claude Opus 4.8 is more balanced, with 74.12\% WDR and 76.47\% CAR, although its JVR remains 5.49\%.
This difference matters when tests filter candidate patches: rejecting an incorrect patch is useful only if the suite also accepts correct patches for the issue.

\parabf{Average success hides incomplete task coverage.}
For GPT-5.6-sol, requiring a correct verdict for every candidate reduces the scores to 47.66\% CT-WDR and 28.59\% CT-CAR.
For example, rejecting nine of ten incorrect patches gives a task substantial WDR credit but fails its complete-task rejection condition, because one sampled failure can still pass the suite.
Thus, a high candidate average can coexist with errors on many tasks, each of which would prevent the suite from acting as a complete validator.
Figure~\ref{fig:leaderboard_views} makes this distinction explicit through the upper-right CAR/WDR and CT-CAR/CT-WDR scatter plots and the lower-right JVR bars.
Even high complete-task acceptance or rejection rates need not overlap on the same issues; JVR additionally requires their intersection with reference recognition.
Among the configurations in Table~\ref{tab:main_results}, Kimi-K3 attains the highest CT-WDR at 49.43\% but only 5.98\% JVR, illustrating why optimizing rejection in isolation does not establish reliable validation.

\parabf{Source differences expose the same imbalance.}
For GPT-5.6-sol, WDR increases from 82.57\% on Multilingual to 96.82\% on DeepSWE, while CAR falls from 79.19\% to 34.26\%.
Its JVR drops from 16.17\% to 2.73\%; the best Pro JVR is 8.77\% with Claude-Fable-5 (Table~\ref{tab:split_leaders}).
Thus, high rejection performance alone does not establish reliable joint validation, especially when many correct patches are rejected.
The source splits differ in issue composition and candidate coverage, so these results characterize the observed difficulty without isolating its cause.

These scores reveal disagreements with the official evaluation; understanding them requires examining which behaviors a test actually checks.

\begin{figure}[!htbp]
\centering
\includegraphics[width=\linewidth]{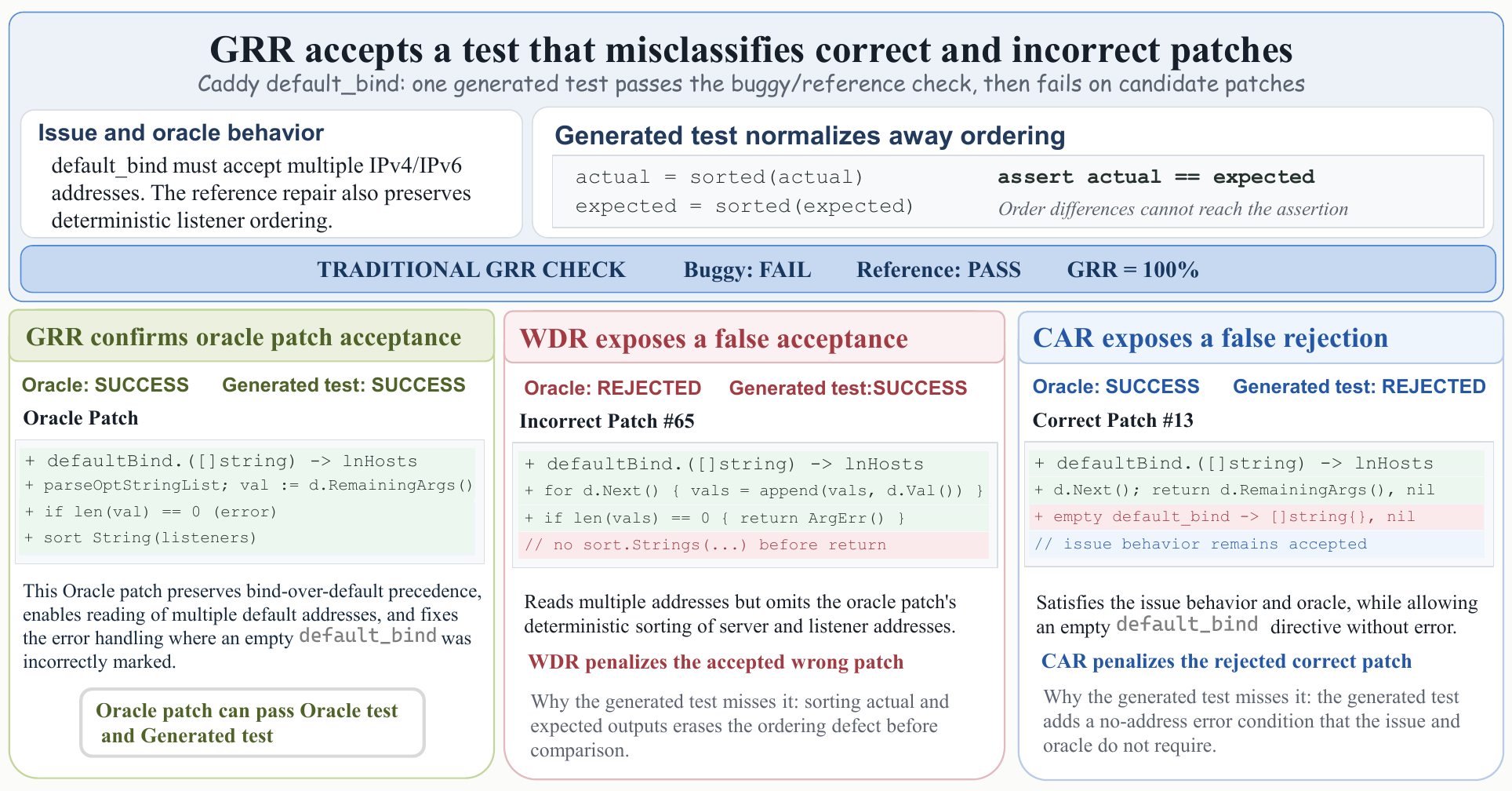}
\caption{Patch-discrimination case study for Caddy's \texttt{default\_bind} option.}
\label{fig:case_study}
\end{figure}

\noindent\begin{minipage}{\linewidth}
\parabf{Case study: the limits of reference reproduction.}
\label{par:case_study}
Figure~\ref{fig:case_study} examines a GPT-5.6-sol test generated with Codex for Caddy instance~4774 from SWE-bench Multilingual.
The issue arises when a server uses \texttt{default\_bind} to specify multiple IPv4/IPv6 addresses: configuration parsing fails on the second address, preventing startup.
The requested fix must support this global setting while allowing individual sites to override it.
\end{minipage}
\par

The test fails on the buggy codebase and passes with the reference patch, yet accepts incorrect patch~\#65 and rejects correct patch~\#13 under the official labels.
Its comparisons ignore address order, which the official tests check.
Conversely, it requires an error for an empty \texttt{default\_bind}, introducing a condition not required by the issue or oracle and thereby rejecting correct patch~\#13.
The issue specifies neither behavior.
This suite thus overlooks one distinction enforced by the official tests while checking another beyond their coverage.
Both choices are compatible with reference success, which therefore establishes neither the rejection of incorrect patches nor the acceptance of correct ones.
Evaluating both candidate classes exposes this gap; interpreting individual disagreements still requires checking the issue requirements and official test coverage.
Across this task's pools, the test rejects 66.2\% of incorrect patches and accepts 49.1\% of correct patches, failing joint validation.
Appendix~\ref{app:audits} gives further audit details and examples.

\begin{summary}
\textbf{Answer to RQ1:} Reference success rarely extends to complete patch
discrimination; evaluating both candidate classes exposes differences in
coverage that the reference transition leaves untested.
\end{summary}

\subsection{RQ2: Selecting Test-generation Demonstrations}
\label{sec:rq2}


All training configurations use Qwen3-30B-A3B-Instruct from the Qwen3 family~\citep{yang2025qwen3technicalreport} for five epochs, with trajectories generated by DeepSeek-V4-Flash. 
All resulting SFT models are evaluated using OpenHands.
The shared SWE SFT baseline is trained on 1.9 billion tokens of repository-level repair trajectories.
We augment this baseline with TestGen trajectories while keeping all repositories used for SFT disjoint from those in \benchmark and the downstream evaluation benchmarks.
Appendix~\ref{app:sft} describes the data composition and evaluation protocol.
We compare three TestGen additions: \textbf{(i) TestGen Supervision}, selected by GRR; \textbf{(ii) Validation-guided TestGen}, selected by WDR/CAR/JVR; and \textbf{(iii) Scaled TestGen Supervision}, which expands the validation-selected data.
We select and train on full TestGen trajectories rather than final test scripts alone.
Table~\ref{tab:sft_results} abbreviates AVG@4 and Pass@4 as AVG and Pass, respectively, and annotates percentage-point changes from SWE SFT above each score. AVG@4 measures the mean per-round resolve rate, while Pass@4 measures the fraction of tasks resolved in at least one of four rounds.

\parabf{Training-data budgets.}
TestGen Supervision and Validation-guided TestGen each add 2 billion tokens of TestGen trajectories to the same SWE SFT baseline, selected from the same trajectory pool using GRR and WDR/CAR/JVR, respectively.
This pool is constructed from approximately 20,000 issue environments.
Scaled TestGen Supervision retains the 2 billion tokens of validation-selected TestGen trajectories and adds another 2 billion tokens selected with the same criteria from an expanded pool, yielding 4 billion tokens of TestGen data in total.

\parabf{Validation-guided TestGen improves quality beyond reproduction.}
We evaluate all four SFT configurations directly on \benchmark using OpenHands and report their downstream repair scores in the same table.
Compared with GRR-selected TestGen Supervision, Validation-guided TestGen increases JVR from 2.26\% to 3.23\% and F1 from 50.39\% to 56.45\%.
GRR increases from 39.26\% to 40.39\%, WDR from 59.00\% to 64.37\%, and CAR from 43.98\% to 50.27\%, showing gains across all three validation dimensions: reproducing the issue, rejecting incorrect patches, and accepting correct patches.
Scaled TestGen Supervision further reaches 3.88\% JVR and 60.18\% F1, compared with 1.29\% and 22.79\% for SWE SFT.
These results suggest that patch-discrimination signals provide a useful basis for selecting TestGen demonstrations beyond GRR-based selection alone.

\begin{table}[htbp]
\centering
\caption{
Test-generation and downstream repair performance across SFT configurations.
}
\label{tab:sft_results}

\begingroup

\footnotesize
\setlength{\tabcolsep}{2pt}
\renewcommand{\arraystretch}{1.85}

\providecommand{\sftlabel}[2]{%
  \raisebox{-0.5pt}{%
    \begingroup
    \small
    \shortstack[l]{#1\\#2}%
    \endgroup
  }%
}

\providecommand{\sftcell}[2]{%
  \makebox[27pt][c]{%
    #1%
    \rlap{%
      \kern-12pt
      \raisebox{10.4pt}{%
        \begingroup
        \fontsize{6}{7}\selectfont
        \textcolor{black!60}{\strut #2}%
        \endgroup
      }%
    }%
  }%
}

\providecommand{\sftredcell}[2]{%
  \makebox[27pt][c]{%
    #1%
    \rlap{%
      \kern-12pt
      \raisebox{10.4pt}{%
        \begingroup
        \setlength{\fboxsep}{1pt}%
        \fontsize{6}{7}\selectfont
        \colorbox{red!10}{%
          \textcolor{red!55!black}{\strut #2}%
        }%
        \endgroup
      }%
    }%
  }%
}

\providecommand{\sftbestcell}[2]{%
  \makebox[27pt][c]{%
    #1%
    \rlap{%
      \kern-12pt
      \raisebox{10.4pt}{%
        \begingroup
        \setlength{\fboxsep}{1pt}%
        \fontsize{6}{7}\selectfont
        \colorbox{green!14}{%
          \textcolor{green!40!black}{\strut #2}%
        }%
        \endgroup
      }%
    }%
  }%
}

\resizebox{\linewidth}{!}{%
\begin{tabular}{
  >{\raggedright\arraybackslash}p{86pt}
  *{14}{c}
  @{\hspace{4pt}}
}

\toprule

\multirow{2}{*}{%
  {\normalsize Configuration}%
}
& \multicolumn{8}{c}{\benchmark}
& \multicolumn{2}{c}{Verified}
& \multicolumn{2}{c}{Multilingual}
& \multicolumn{2}{c}{Pro}
\\[-2pt]

\cmidrule(lr){2-9}
\cmidrule(lr){10-11}
\cmidrule(lr){12-13}
\cmidrule(l){14-15}

&
GRR
& WDR
& CAR
& F1
& CT-WDR
& CT-CAR
& CT-F1
& JVR
& AVG
& Pass
& AVG
& Pass
& AVG
& Pass
\\[-1pt]

\midrule

{\small SWE SFT}
& 8.89
& 20.75
& 25.28
& 22.79
& 9.21
& 19.39
& 12.49
& 1.29
& 67.53
& 79.20
& 55.08
& 72.00
& 46.56
& 60.05
\\

\addlinespace[3pt]

\sftlabel{+ TestGen}{Supervision}
& \sftcell{39.26}{+30.37}
& \sftcell{59.00}{+38.25}
& \sftcell{43.98}{+18.70}
& \sftcell{50.39}{+27.60}
& \sftcell{13.73}{+4.52}
& \sftcell{28.11}{+8.72}
& \sftcell{18.45}{+5.96}
& \sftcell{2.26}{+0.97}
& \sftcell{69.47}{+1.94}
& \sftcell{80.00}{+0.80}
& \sftcell{62.05}{+6.97}
& \sftcell{76.25}{+4.25}
& \sftcell{48.97}{+2.41}
& \sftcell{60.47}{+0.42}
\\

\addlinespace[3pt]

\sftlabel{+ Validation-guided}{TestGen}
& \sftredcell{40.39}{+31.50}
& \sftredcell{64.37}{+43.62}
& \sftredcell{50.27}{+24.99}
& \sftredcell{56.45}{+33.66}
& \sftredcell{25.69}{+16.48}
& \sftredcell{31.50}{+12.11}
& \sftredcell{28.30}{+15.81}
& \sftredcell{3.23}{+1.94}
& \sftredcell{70.24}{+2.71}
& \sftredcell{80.60}{+1.40}
& \sftredcell{63.02}{+7.94}
& \sftredcell{76.33}{+4.33}
& \sftredcell{49.18}{+2.62}
& \sftredcell{60.68}{+0.63}
\\

\addlinespace[3pt]

\sftlabel{+ Scaled TestGen}{Supervision}
& \sftbestcell{\textbf{41.20}}{+32.31}
& \sftbestcell{\textbf{72.14}}{+51.39}
& \sftbestcell{\textbf{51.62}}{+26.34}
& \sftbestcell{\textbf{60.18}}{+37.39}
& \sftbestcell{\textbf{31.34}}{+22.13}
& \sftbestcell{\textbf{32.96}}{+13.57}
& \sftbestcell{\textbf{32.13}}{+19.64}
& \sftbestcell{\textbf{3.88}}{+2.59}
& \sftbestcell{\textbf{71.84}}{+4.31}
& \sftbestcell{\textbf{82.00}}{+2.80}
& \sftbestcell{\textbf{63.30}}{+8.22}
& \sftbestcell{\textbf{78.00}}{+6.00}
& \sftbestcell{\textbf{50.70}}{+4.14}
& \sftbestcell{\textbf{61.70}}{+1.65}
\\[1pt]

\bottomrule

\end{tabular}%
}

\endgroup
\end{table}

\parabf{Selection improves complete-task coverage.}
At the same TestGen budget, validation-guided selection raises CT-WDR from 13.73\% to 25.69\% and CT-CAR from 28.11\% to 31.50\% relative to GRR-based selection.
More tasks therefore receive correct verdicts for every sampled incorrect patch, while acceptance of the complete correct pool also improves.
Scaling raises these rates further to 31.34\% and 32.96\%, respectively.
These gains extend beyond reference recognition, although satisfying both conditions together with reproduction remains difficult.

\parabf{Interpreting the training comparison.}
Paired tests quantify the gains over SWE SFT (Appendix~\ref{app:statistical_reliability_rq2}).
The 95\% bootstrap intervals for JVR improvements are $[0.81,3.23]$ percentage points for Validation-guided TestGen and $[1.29,4.04]$ for Scaled TestGen Supervision.
Both improvements are significant under exact two-sided McNemar tests with Holm correction over the three baseline comparisons ($p=.008$ and $.001$, respectively), whereas the GRR-selected comparison is not ($p=.146$).
These comparisons establish gains relative to SWE SFT; they do not directly test the difference between the two selection recipes.

\begin{summary}
\textbf{Answer to RQ2:} At the same data budget, validation-guided selection outperforms GRR-based selection, with further gains from scaling the selected trajectories.

\end{summary}

\subsection{RQ3: Transfer to Downstream Repair}

\parabf{TestGen supervision improves repair success.}
To test whether the training benefits extend beyond standalone test generation, we evaluate repository repair with OpenHands on SWE-bench Verified, Multilingual, and Pro using AVG@4 and Pass@4.
Adding GRR-selected TestGen Supervision raises AVG@4 over SWE SFT by 1.94, 6.97, and 2.41 percentage points on Verified, Multilingual, and Pro, respectively.
Validation-guided TestGen further improves AVG@4 by 0.77 percentage points on Verified, 0.97 on Multilingual, and 0.21 on Pro.
Pass@4 also increases on all three benchmarks.

Scaled TestGen Supervision further reaches AVG@4/Pass@4 of 71.84\%/82.00\% on Verified, 63.30\%/78.00\% on Multilingual, and 50.70\%/61.70\% on Pro.
Its AVG@4 gains over SWE SFT are 4.31, 8.22, and 4.14 percentage points, respectively.
Together with the improvements in patch-discrimination metrics in RQ2, these results show that scaling validation-guided TestGen supervision benefits both standalone test generation and downstream repair.

\noindent\begin{minipage}{\linewidth}
\parabf{Scaled supervision also shortens interactions.}
Mean interaction length decreases from 106.68 to 80.70 steps on Verified, 157.62 to 115.70 on Multilingual, and 151.86 to 119.22 on Pro.
These reductions accompany higher repair success, although interaction length does not directly measure elapsed time or token cost.

\begin{summary}
\textbf{Answer to RQ3:} Scaled TestGen Supervision improves repair success and task coverage across all three downstream benchmarks with fewer interaction steps.
\end{summary}
\end{minipage}

\section{Related Work}

\parabf{Generating tests for existing code and reported issues.}
TestEval, TestGenEval, CLOVER, and MultiFileTest assess test generation across code granularities~\citep{wang-etal-2025-testeval,jain2025testgenevalrealworldunit,xu2025clovertestcasegeneration,wang-etal-2026-multifiletest}.
SWE-Compass evaluates patch-conditioned test generation for correctness and coverage~\citep{xu2025swecompassunifiedevaluationagentic}.
TestGen-LLM filters test extensions for build success, reliable execution, and coverage gains~\citep{Alshahwan_2024}.
CoverUp and TestForge refine tests using coverage and execution feedback~\citep{Altmayer_Pizzorno_2025,jain2025testforgefeedbackdrivenagentictest}.
Coverage alone need not imply effective fault detection~\citep{Inozemtseva_2014}.
LIBRO, AEGIS, and Otter use issues to guide reproduction or validate changes~\citep{Kang_2023,wang2024aegisagentbasedframeworkgeneral,pmlr-v267-ahmed25b}.

\parabf{Using generated tests to assess candidates.}
CodeT selects solutions to programming tasks using generated tests and execution agreement~\citep{chen2022codetcodegenerationgenerated}.
DiffTGen uses generated tests and an oracle to detect test-suite-overfitted patches~\citep{Xin_2017}.
SWT-Bench and TDD-Bench Verified evaluate repository tests through buggy-to-reference fail-to-pass transitions~\citep{NEURIPS2024_94f093b4,ahmed2024tddbenchverifiedllmsgenerate}.
BRT Agent ranks candidate patches by their pass rates over generated tests~\citep{cheng2025agenticbugreproductioneffective}.
Our protocol fixes correct and incorrect patch pools across configurations, measuring acceptance, rejection, and reproduction jointly.

\parabf{Assessing tests through program changes.}
Mutation testing measures whether tests detect injected faults~\citep{DeMillo_1978,ORBi-aaa51123-f7a7-4be2-a56a-a19ececa5f00}.
Mutant detection correlates with real-fault detection, although available mutation operators leave some faults unrepresented~\citep{Just_2014}.
OmniCode uses incorrect patches from agent attempts and LLM perturbations~\citep{sonwane-etal-2026-omnicode}, while SWE-Mutation generates agentic mutants for 800 instances spanning ten languages~\citep{sun2026swemutationllmsgeneratereliable}.
Our pools include correct patches, so evaluation penalizes their rejection as well as missed incorrect patches.
Execution-based selection, trajectory supervision, and scaled repository data support training code models and tool-use agents~\citep{NEURIPS2024_72da102d,pmlr-v267-pan25g,NEURIPS2025_8b86cf5a,zhan2025katcodertechnicalreport,tian2026sapstateguideddatasynthesis}.
Our acceptance and rejection signals guide demonstration selection for test generation and downstream repair.

\section{Conclusion}
\label{sec:conclusion}

We introduced \benchmark to evaluate whether generated suites reproduce issues, accept correct patches, and reject incorrect patches across 619 tasks and 82,579 candidate patches.
The results show that reference recognition and high scores on individual candidate classes do not ensure complete validation.
Validation-guided selection and scaled TestGen supervision improve both test generation and downstream repair, making patch discrimination useful for benchmarking and training-data selection.

\parabf{Scope of the findings.}
Candidate labels follow the official oracle, and JVR evaluates joint validation over the released candidate pools rather than every possible patch.
The reproduction-filtered challenge set targets difficult issues and does not represent all repository tasks.
Bootstrap intervals quantify task-sampling uncertainty conditional on the recorded runs and candidate pools.
Appendix~\ref{app:limitations} details these boundaries.
Auditing these boundaries can guide training toward tests that generalize across implementations.

\label{sec:main_end}

\clearpage
\subsection*{AI use statement}
\label{sec:ai_use}

LLM-based coding agents generated candidate patches and test-generation trajectories, and an LLM judge assessed reference-test quality for task selection, as described in Section~\ref{sec:benchmark} and Appendix~\ref{app:sft}.
Codex assisted with refining the evaluation framework and metrics, interpreting quantitative results and qualitative patch examples, locating literature, revising manuscript structure and language, and preparing LaTeX tables and plotting code.
AI-assisted numerical summaries and tables were checked against the recorded results, and the manuscript was compiled and checked for numerical and reference consistency.
The authors take responsibility for the final text, claims, data, and artifacts, including content produced with AI assistance.

\bibliographystyle{iclr2027_conference}
\bibliography{custom,iclr2027_conference}

\clearpage
\appendix
\raggedbottom

\section{Task Representation and Curation}
\label{app:voting}
\label{app:task_curation}

\subsection{Task Record and Evidence}

We represent each task by its issue, program state, and evaluation evidence:
\begin{equation}
\mathcal{D}_i=(I_i,R_i,p_i^{ref},X_i,H_i,P_i^{ok},P_i^{bad}).
\end{equation}
Here, $I_i$ is the issue description, $R_i$ the buggy codebase, $p_i^{ref}$ the reference patch, $X_i$ the official test patch, and $H_i$ the original execution harness.
Applying a patch produces the codebase $R_i\circ p$, following the notation in Section~\ref{sec:protocol}.
The sets $P_i^{ok}$ and $P_i^{bad}$ contain correct and incorrect patches, respectively, excluding the separately evaluated reference patch.
As in Section~\ref{sec:benchmark}, a correct label records harness acceptance rather than complete coverage of the issue specification.

The record must support three complementary checks, because a successful reference transition alone does not establish useful candidate diversity.
\textbf{(i) Reference quality:} the official tests establish a meaningful buggy-to-reference transition.
\textbf{(ii) Correct-patch diversity:} correct patches allow tests to be evaluated across different implementations.
\textbf{(iii) Incorrect-patch diversity:} incorrect patches expose plausible ways to leave issue requirements unsatisfied.

\subsection{Reference-Test Quality}



We apply the same LLM-based audit across all three benchmark sources.
For each task, the audit jointly examines the issue description, reference patch, and official tests, and assigns a quality score from 0 to 100 based on assertion strength, avoidance of implementation-specific checks, whether the tests fail on the buggy program, and relevance to the reported issue.
Using a threshold of 80, the quality audit retains 461 SWE-bench Pro tasks, 278 SWE-bench Multilingual tasks, and 110 DeepSWE tasks.
We additionally manually review the audit outcomes before finalizing task eligibility.
Together, the LLM-based audit and human review provide quality control over the official tests used as the execution oracle for candidate verification.

The quality audit determines task eligibility only; candidate-patch labels and generated-suite scores are subsequently determined through execution.

\subsection{Reference-Transition Difficulty}

We further filter tasks using a heterogeneous panel of test-generation configurations, removing tasks whose reference fail-to-pass transition is consistently reproduced across the panel.
This filtering step removes tasks that are broadly easy for existing test-generation systems and retains tasks with nontrivial reference-transition difficulty.

After both curation stages, the benchmark retains 342 SWE-bench Pro tasks,
167 SWE-bench Multilingual tasks, and 110 DeepSWE tasks, for a total of 619 tasks.
Table~\ref{tab:filter_funnel} summarizes the resulting task counts by source.

\begin{table}[H]
\centering
\caption{Task-level curation by benchmark source.}
\label{tab:filter_funnel}

\begingroup
\small
\setlength{\tabcolsep}{8pt}
\renewcommand{\arraystretch}{1.12}

\begin{tabular}{@{\hspace{8pt}}lrrr@{\hspace{8pt}}}
\toprule
Source & Initial & Removed & Retained (\%) \\
\midrule
SWE-bench Pro          & 731 & 389 & 342 \;(46.8) \\
SWE-bench Multilingual & 300 & 133 & 167 \;(55.7) \\
DeepSWE                & 113 &   3 & 110 \;(97.3) \\
\midrule
\textbf{Overall}
& \textbf{1,144}
& \textbf{525}
& \textbf{619 \;(54.1)} \\
\bottomrule
\end{tabular}

\endgroup
\end{table}

\FloatBarrier

\clearpage
\subsection{Task-Curation Judge Prompt}
\label{app:task_curation_prompt}

We use an LLM-based quality audit to determine whether each candidate
issue provides sufficiently reliable evaluation evidence for inclusion
in \benchmark.
For each issue, the judge examines the issue description, reference
patch, and official test patch.
The audit evaluates whether the required behavior is objectively
testable, whether the official tests meaningfully enforce that behavior,
whether the assertions are sufficiently strong, whether the expected
buggy-to-repaired transition is supported, and whether the tests
remain independent of the particular implementation in the reference
patch.
The resulting recommendation is \texttt{include}, \texttt{review}, or
\texttt{exclude}; cases with incomplete or ambiguous evidence are
referred for manual review.

Figure~\ref{fig:task_curation_prompt} presents the task-curation prompt
used to organize these judgments.
The displayed version condenses repeated explanations while preserving
the judge's inputs, evaluation criteria, decision policy, and output
requirements.

\begin{figure}[H]
\centering
\begin{tcolorbox}[
  colback=white,
  colframe=black!45,
  colbacktitle=black!8,
  coltitle=black,
  boxrule=0.4pt,
  arc=1pt,
  left=7pt,right=7pt,top=5pt,bottom=5pt,
  fonttitle=\sffamily\bfseries\small,
  fontupper=\sffamily\fontsize{8}{9.2}\selectfont,
  title={System Prompt for Task Curation}
]
\setlength{\parindent}{0pt}
\setlength{\parskip}{2pt}

{\centering\bfseries - ROLE \& OBJECTIVE -\par}
You are a strict senior software engineer auditing candidate issues for inclusion in a software-engineering benchmark.
Given an issue description, a reference patch, and an official test patch, determine whether the task provides a reliable and implementation-independent evaluation oracle.

Your goal: retain tasks whose tests meaningfully enforce the reported behavior while allowing valid alternative repairs.
The reference patch is only one correct implementation; judge externally observable behavior, not internal code structure.
When evidence is genuinely ambiguous, request human review.

\textbf{Input evidence:}\par
- Issue / problem statement: required behavior and scope.\par
- Reference patch: one official implementation of the repair.\par
- Official test patch: assertions intended to distinguish buggy from correctly repaired behavior.

{\centering\bfseries - EVALUATION CRITERIA -\par}
\textbf{A. Issue clarity and alignment.} Is the required behavior objectively testable?
Does the reference patch address it, and do the official tests directly exercise and assert that behavior rather than unrelated effects?

\textbf{B. Assertion strength.} Could an incomplete or incorrect repair still pass?
Flag tests that only check execution, nonempty output, broad types, incidental examples, setup behavior, or permissive conditions.

\textbf{C. Implementation independence.} Would another valid repair fail because the test encodes private helpers, exact call sequences, internal control flow, patch location, incidental logs or ordering, source matching, or copied constants?

\textbf{D. Reference-transition reliability.} Does the evidence support tests(Buggy) = fail and tests(Correct repair) = pass?
Separate behavioral failures from compilation, dependency, timeout, flaky, external-service, or infrastructure failures.

{\centering\bfseries - DECISION POLICY -\par}
\textbf{1. INCLUDE:} clear, testable behavior; aligned and meaningful assertions; sufficient strength; alternative valid implementations remain admissible.

\textbf{2. REVIEW:} evidence is incomplete or genuinely ambiguous, or the judgment depends on repository context absent from the supplied artifacts.

\textbf{3. EXCLUDE:} issue-test mismatch, materially weak tests, reference-patch overfitting, unsupported fail-to-pass transition, or no objectively assessable requirement.

{\centering\bfseries - OUTPUT FORMAT -\par}
Return only one JSON object containing: an evidence-based summary; required and tested behavior; Boolean judgments for clarity, alignment, strength, implementation independence, and transition support; concrete evidence for weak tests, overfitting, or unrelated-failure risk; 0{-}{-}100 component scores; decision = include \textbar{} review \textbar{} exclude; a primary reason; and confidence in [0, 1].

\textbf{Task-specific fields:}\par
\texttt{ISSUE\_DESCRIPTION}\par
\texttt{REFERENCE\_PATCH}\par
\texttt{OFFICIAL\_TEST\_PATCH}

\textbf{Final instruction:} Evaluate whether this task provides sufficiently reliable and implementation-independent test evidence for benchmark inclusion.
Return only the required JSON object.
\end{tcolorbox}
\caption{Simplified system prompt for the task-curation judge.}
\label{fig:task_curation_prompt}
\end{figure}

\FloatBarrier

\section{Candidate-Pool Construction}
\label{app:pools}

\parabf{Patch producers and model roles.}
Table~\ref{tab:collection_models} lists the model inventory used to collect candidate patches.
Candidate generation, task-selection voting, and the main test-generation comparison are separate uses of these models; a model's appearance in the inventory does not imply that it participates in every stage.

\begin{table}[!htbp]
\centering
\caption{
Model families used for candidate patch collection.
}
\label{tab:collection_models}

\begingroup
\small
\setlength{\tabcolsep}{6pt}
\renewcommand{\arraystretch}{1.14}

\begin{tabularx}{\textwidth}{
  @{}
  >{\raggedright\arraybackslash\bfseries}p{0.13\textwidth}
  >{\raggedright\arraybackslash}X
  @{}
}
\toprule
Model family
&
Model configurations
\\
\midrule

Claude
&
\mbox{Claude-Fable-5.1},
\mbox{Claude-Opus-5},
\mbox{Claude-Sonnet-5},
\mbox{Claude-Fable-5},
\mbox{Claude-Opus-4.8}
\\
\addlinespace[0.35em]

GPT
&
\mbox{GPT-6-Astra},
\mbox{GPT-5.6-Sol},
\mbox{GPT-5.4},
\mbox{GPT-5.2}
\\
\addlinespace[0.35em]

DeepSeek
&
\mbox{DeepSeek-V4-Pro},
\mbox{DeepSeek-V4-Flash}
\\
\addlinespace[0.35em]

Gemini
&
\mbox{Gemini-3.8-Flash},
\mbox{Gemini-3.1-Pro},
\mbox{Gemini-3-Pro},
\mbox{Gemini-2.5-Pro},
\mbox{Gemini-2.5-Flash}
\\
\addlinespace[0.35em]

GLM
&
\mbox{GLM-5.3},
\mbox{GLM-5.2}
\\
\addlinespace[0.35em]

Qwen
&
\mbox{Qwen3.8-Max},
\mbox{Qwen3.8-27B},
\mbox{Qwen3.6-35B-A3B},
\mbox{Qwen3.5-35B-A3B}
\\
\addlinespace[0.35em]

Kimi
&
\mbox{Kimi-K3}
\\

\bottomrule
\end{tabularx}

\endgroup
\end{table}

\parabf{Normalizing diffs.}
Deduplication operates independently within each task and correctness class.
We strip hunk line numbers, index lines, trailing whitespace, and unchanged context, retaining the added and deleted lines.
This normalization reduces differences caused solely by diff formatting.

\parabf{Filtering similar candidates.}
We filter the normalized diffs at character-level TF--IDF similarity $\tau=0.9$.
Inverse-document-frequency weighting downweights features shared across many patches~\citep{SPARCK_JONES_1972}.
The feature representation uses character $n$-grams with $n\in\{3,4,5\}$.
We process diffs from longest to shortest, discarding a candidate if its cosine similarity to any retained diff is at least $0.9$.
Character features avoid language-specific tokenizers, although textual distance remains an approximation to implementation diversity.

\parabf{Capping each class.}
If more than 128 candidates remain in a class, farthest-point sampling selects 128 spread across the feature space.
Both classes use the same cap; the larger incorrect pool reflects the greater number of distinct incorrect patches retained.
Final curation removes candidates overlapping official test hunks and entries without a usable solution artifact.
The final pools contain 82,579 candidates: 27,749 correct and 54,830 incorrect patches.
Table~\ref{tab:pool_stages} reports their counts and per-task means over all 619 tasks.

\begin{table}[htbp]
\centering
\caption{Candidate-pool composition by correctness class.}
\label{tab:pool_stages}

\begingroup
\small
\setlength{\tabcolsep}{12pt}
\renewcommand{\arraystretch}{1.15}

\begin{tabular}{@{\hspace{8pt}}lrr@{\hspace{8pt}}}
\toprule
Patch label & Count & Average per task \\
\midrule
Correct patches   & 27,749 & 44.83 \\
Incorrect patches & 54,830 & 88.58 \\
\textbf{All candidate patches}
& \textbf{82,579}
& \textbf{133.41} \\
\bottomrule
\end{tabular}

\endgroup
\end{table}

\FloatBarrier

\section{Evaluation Protocol and Metrics}

\subsection{Generated-Suite Protocol}
\label{app:suite}

\parabf{Extracting and isolating artifacts.}
The submitted package consists of a test script, fixtures, and supporting files in a task-local \texttt{tmp} directory, without changes to existing repository files.
Separating this package from the official test patch avoids conflicts that would confound applicability with test quality.
We extract the script and referenced files directly from \texttt{tmp}, rather than reconstructing the package from a Git diff.
Identical packages are installed in clean copies of the buggy codebase and codebases obtained by applying the reference or a candidate patch.
We reset between executions so that fixtures, caches, and side effects cannot carry over to the next program state.

\parabf{Required outcomes.}
Using the execution function defined in Section~\ref{sec:protocol}, the desired suite outcomes are
\begin{equation}
\begin{aligned}
\operatorname{exec}(T_i,R_i)&=\mathrm{F}, &
\operatorname{exec}(T_i,R_i\circ p_i^{ref})&=\mathrm{P},\\
\operatorname{exec}(T_i,R_i\circ p)&=\mathrm{P}\quad(p\in P_i^{ok}), &
\operatorname{exec}(T_i,R_i\circ p)&=\mathrm{F}\quad(p\in P_i^{bad}).
\end{aligned}
\end{equation}
Suite-level acceptance requires every submitted test to pass.
Individual tests can also be replayed to identify the assertions responsible for a suite verdict.

\parabf{Recording outcomes.}
For each suite execution, logs record the command, exit code, timeout state, failing assertions, and harness output.
Outcomes distinguish completed passes, stable behavioral failures, timeouts, infrastructure failures, and missing output.
We replay inconsistent outcomes and withhold both acceptance and rejection credit if instability persists.
Flaky tests can produce different outcomes for unchanged code~\citep{Luo_2014}; execution statuses must therefore be examined before assigning a behavioral explanation.

\subsection{Metric Definitions and Scoring Conventions}
\label{app:metrics}

Every benchmark rate retains the same denominator of $N=619$ tasks.
Equation~\ref{eq:execution_indicators} defines the indicators from $\operatorname{exec}$: $a_i(p)=1$ when the suite passes on $R_i\circ p$, and $r_i(p)=1$ when a completed run produces a stable behavioral failure.
Unresolved executions, missing artifacts, and infrastructure-only outcomes are recorded as $\bot$ and set both indicators to zero.
Within-task denominators use the released pool sizes, preventing failed executions from improving a score by disappearing from its denominator.

\parabf{Reference recognition: GRR.}
Gold Recognition Rate averages the reference fail-to-pass indicator over tasks:
\begin{equation}
\begin{aligned}
g_i&=\mathbf{1}[\operatorname{exec}(T_i,R_i)=\mathrm{F}\ \wedge\ \operatorname{exec}(T_i,R_i\circ p_i^{ref})=\mathrm{P}],\\
\mathrm{GRR}&=\frac{1}{N}\sum_{i=1}^{N}g_i.
\end{aligned}
\end{equation}
We set $g_i=0$ when the suite or either required transition outcome is unavailable.
This check establishes reference reproduction without imposing verdicts on candidate patches.

\parabf{Candidate discrimination: WDR and CAR.}
For each nonempty pool, we compute the fraction of incorrect patches rejected or correct patches accepted:
\begin{equation}
w_i=\frac{\sum_{p\in P_i^{bad}}r_i(p)}{|P_i^{bad}|},
\qquad
c_i=\frac{\sum_{p\in P_i^{ok}}a_i(p)}{|P_i^{ok}|}.
\end{equation}
WDR and CAR are the macro averages of these within-task fractions:
\begin{equation}
\mathrm{WDR}=\frac{1}{N}\sum_{i=1}^{N}w_i,
\qquad
\mathrm{CAR}=\frac{1}{N}\sum_{i=1}^{N}c_i.
\end{equation}
An issue contributes the same weight regardless of how many candidates it contains.

WDR evaluates rejection of incorrect patches, whereas CAR evaluates preservation of correct patches.
A suite that rejects all candidates can score highly on WDR while failing CAR, illustrating why either rate alone is insufficient.
Execution failures also lower these rates, so behavioral diagnoses must be checked against the recorded statuses.

\parabf{Complete-task discrimination: CT-WDR, CT-CAR, and JVR.}
Complete-task rates count only issues on which the suite rejects every incorrect patch or accepts every correct patch:
\begin{equation}
\mathrm{CT\text{-}WDR}=\frac{1}{N}\sum_{i=1}^{N}\mathbf{1}[w_i=1],
\qquad
\mathrm{CT\text{-}CAR}=\frac{1}{N}\sum_{i=1}^{N}\mathbf{1}[c_i=1].
\label{eq:complete_task}
\end{equation}
JVR further requires reproduction and both complete-task conditions to hold together:
\begin{equation}
\mathrm{JVR}=\frac{1}{N}\sum_{i=1}^{N}\mathbf{1}[g_i=1\ \wedge\ w_i=1\ \wedge\ c_i=1].
\end{equation}
JVR determines the reported ranking, while CT-WDR and CT-CAR identify the class that prevents complete validation.
Complete-task comparisons keep the pools fixed because adding or diversifying candidates can introduce further opportunities for failure.

\parabf{Balanced summaries: F1 and CT-F1.}
The two balanced summaries use harmonic means of the corresponding rates:
\begin{equation}
\begin{aligned}
\mathrm{F1}&=h(\mathrm{WDR},\mathrm{CAR}), &
\mathrm{CT\text{-}F1}&=h(\mathrm{CT\text{-}WDR},\mathrm{CT\text{-}CAR}),\\
h(a,b)&=\frac{2ab}{a+b}\quad(a+b>0), & h(0,0)&=0.
\end{aligned}
\label{eq:harmonic}
\end{equation}
These summaries combine benchmark rates rather than pooled classifier precision and recall.
They reward aggregate acceptance--rejection balance but do not impose the task-level conjunction measured by JVR.

\subsection{Agent Configurations}
\label{app:agent_configurations}
The main benchmark evaluates each model with its native coding agent where available: GPT-5.6-sol uses Codex; the Claude Fable and Opus models use Claude Code; Kimi-K3 uses Kimi Code; Qwen3.8-max uses Qwen Code; and GLM-5.3 uses ZCode.
All four DeepSeek-V4 configurations, including Pro, Flash, and their Preview variants, use OpenHands.
These agents generate test packages under the common artifact and execution protocol above, and all configurations are scored with the same benchmark metrics.
The four direct SFT evaluations and downstream repair evaluations both use OpenHands.
Figure~\ref{fig:leaderboard_views} shortens GPT-5.6-sol to GPT-5.6, Claude models to Fable/Opus, DeepSeek-V4 to DS, Preview to prv, and Qwen3.8-max to Qwen3.8.

\FloatBarrier

\section{RQ1: Additional Results and Analysis}
\label{app:rq1}

\subsection{Benchmark Leaderboard Reliability}
\label{app:statistical_reliability_rq1}

\label{app:statistical_reliability}

\parabf{Resampling protocol.}
We quantify uncertainty over the benchmark task set using a paired,
source-stratified nonparametric bootstrap~\citep{Efron_1979}.
Each of the 10,000 bootstrap replicates samples tasks with replacement
within the three benchmark sources: 342 SWE-bench Pro tasks,
167 SWE-bench Multilingual tasks, and 110 DeepSWE tasks.
The same sampled task indices are applied to all configurations, preserving
both the source composition and the paired structure of model comparisons.
For every replicate, we recompute GRR, WDR, CAR, CT-WDR, CT-CAR, and JVR
from the sampled task-level outcomes.
F1 and CT-F1 are subsequently recomputed as the harmonic means of
WDR and CAR and of CT-WDR and CT-CAR, respectively.
We report two-sided percentile 95\% confidence intervals.

The RQ1 analysis covers the eight representative coding-agent configurations
included in the main benchmark comparison
(Table~\ref{tab:main_results}).
Table~\ref{tab:bootstrap_rq1_metrics} reports uncertainty intervals for
all eight benchmark metrics.
Each cell gives the point estimate and its 95\% confidence interval as percentages.
Point estimates match Table~\ref{tab:main_results}; intervals use unrounded task-level values.

\parabf{Paired JVR comparisons.}
We compare each representative configuration with GPT-5.6-sol,
the point-estimate leader on JVR.
For each competitor, $\Delta\mathrm{JVR}$ denotes the competitor's JVR
minus that of GPT-5.6-sol in percentage points under the paired stratified bootstrap.
Because JVR is binary at the task level, we additionally apply an exact
two-sided McNemar test to the discordant task outcomes~\citep{Fagerland2013}.
The seven resulting $p$-values are adjusted using Holm's procedure.
Table~\ref{tab:bootstrap_rq1_jvr} lists discordant task counts as GPT-5.6-sol-only and competitor-only JVR successes, respectively.
Bootstrap rank stability is computed in every replicate; ties for first
divide the corresponding first-place probability equally.

GPT-5.6-sol has the highest observed JVR at 9.21\% and receives
84.5\% of the bootstrap first-place probability.
Claude-Fable-5.1 is the closest configuration, with a JVR difference
of $-1.29$ percentage points and a 95\% confidence interval of
$[-3.72,+0.97]$.
The paired interval and rank distribution therefore provide additional
context for the point-estimate ordering in
Table~\ref{tab:main_results}.

\begingroup
\setlength{\intextsep}{8pt}
\begin{table}[H]
\centering
\caption{
Bootstrap estimates and 95\% confidence intervals for the eight RQ1 coding-agent configurations.
}
\label{tab:bootstrap_rq1_metrics}
\begingroup
\fontsize{9}{10.5}\selectfont
\setlength{\tabcolsep}{3pt}
\renewcommand{\arraystretch}{1.0}
\newcommand{\bootstrapcell}[3]{%
  \begin{tabular}[c]{@{}c@{}}#1\\[-1pt]{\fontsize{8}{9.5}\selectfont [#2, #3]}\end{tabular}%
}

\begin{tabular*}{\linewidth}{@{}l@{\extracolsep{\fill}}cccc@{}}
\toprule
\multicolumn{5}{@{}l}{\textbf{Reference and patch-level metrics}} \\[3pt]
Model & GRR & WDR & CAR & F1 \\
\midrule
GPT-5.6-sol
& \bootstrapcell{59.29}{55.74}{62.84}
& \bootstrapcell{88.57}{86.59}{90.44}
& \bootstrapcell{50.20}{47.20}{53.20}
& \bootstrapcell{64.08}{61.50}{66.54} \\
Claude-Fable-5.1
& \bootstrapcell{65.11}{61.55}{68.66}
& \bootstrapcell{83.74}{81.59}{85.82}
& \bootstrapcell{62.19}{59.14}{65.18}
& \bootstrapcell{71.37}{69.14}{73.51} \\
Claude-Opus-5
& \bootstrapcell{68.17}{64.62}{71.57}
& \bootstrapcell{85.91}{84.15}{87.59}
& \bootstrapcell{66.76}{63.96}{69.54}
& \bootstrapcell{75.13}{73.34}{76.90} \\
Kimi-K3
& \bootstrapcell{33.12}{29.56}{36.51}
& \bootstrapcell{80.24}{77.47}{82.95}
& \bootstrapcell{40.64}{37.33}{43.87}
& \bootstrapcell{53.95}{50.91}{56.80} \\
Qwen3.8-max
& \bootstrapcell{49.60}{45.88}{53.31}
& \bootstrapcell{84.69}{82.81}{86.46}
& \bootstrapcell{61.68}{58.70}{64.65}
& \bootstrapcell{71.38}{69.38}{73.25} \\
DeepSeek-V4-Pro
& \bootstrapcell{36.03}{32.31}{39.90}
& \bootstrapcell{82.53}{80.29}{84.74}
& \bootstrapcell{49.36}{45.91}{52.78}
& \bootstrapcell{61.77}{58.97}{64.39} \\
DeepSeek-V4-Flash
& \bootstrapcell{29.89}{26.33}{33.45}
& \bootstrapcell{80.53}{77.93}{82.98}
& \bootstrapcell{44.62}{41.31}{47.98}
& \bootstrapcell{57.42}{54.55}{60.17} \\
GLM-5.3
& \bootstrapcell{38.77}{34.89}{42.65}
& \bootstrapcell{85.36}{83.38}{87.32}
& \bootstrapcell{53.55}{50.35}{56.70}
& \bootstrapcell{65.81}{63.38}{68.17} \\
\midrule
\addlinespace[3pt]
\multicolumn{5}{@{}l}{\textbf{Complete-task metrics}} \\[3pt]
Model & CT-WDR & CT-CAR & CT-F1 & JVR \\
\midrule
GPT-5.6-sol
& \bootstrapcell{47.66}{43.78}{51.53}
& \bootstrapcell{28.59}{25.20}{31.99}
& \bootstrapcell{35.74}{32.77}{38.52}
& \bootstrapcell{9.21}{6.95}{11.47} \\
Claude-Fable-5.1
& \bootstrapcell{34.89}{31.34}{38.45}
& \bootstrapcell{30.37}{26.82}{33.93}
& \bootstrapcell{32.47}{29.86}{34.91}
& \bootstrapcell{7.92}{5.82}{10.02} \\
Claude-Opus-5
& \bootstrapcell{32.47}{28.92}{36.03}
& \bootstrapcell{31.34}{27.79}{35.06}
& \bootstrapcell{31.89}{29.40}{34.37}
& \bootstrapcell{7.11}{5.17}{9.21} \\
Kimi-K3
& \bootstrapcell{49.43}{45.40}{53.47}
& \bootstrapcell{22.94}{19.71}{26.17}
& \bootstrapcell{31.34}{28.16}{34.26}
& \bootstrapcell{5.98}{4.20}{7.92} \\
Qwen3.8-max
& \bootstrapcell{32.47}{28.92}{36.03}
& \bootstrapcell{30.21}{26.66}{33.76}
& \bootstrapcell{31.30}{28.84}{33.60}
& \bootstrapcell{5.17}{3.55}{6.95} \\
DeepSeek-V4-Pro
& \bootstrapcell{36.35}{32.63}{40.06}
& \bootstrapcell{27.95}{24.56}{31.34}
& \bootstrapcell{31.60}{29.08}{33.94}
& \bootstrapcell{4.52}{3.07}{6.14} \\
DeepSeek-V4-Flash
& \bootstrapcell{40.87}{37.00}{44.75}
& \bootstrapcell{22.94}{19.71}{26.17}
& \bootstrapcell{29.39}{26.52}{32.10}
& \bootstrapcell{3.88}{2.42}{5.49} \\
GLM-5.3
& \bootstrapcell{37.64}{33.93}{41.36}
& \bootstrapcell{25.04}{21.65}{28.43}
& \bootstrapcell{30.07}{27.42}{32.58}
& \bootstrapcell{3.72}{2.42}{5.33} \\
\bottomrule
\end{tabular*}
\endgroup
\end{table}

\begin{table}[H]
\centering
\caption{
Paired JVR comparisons and bootstrap rank stability for RQ1.
}
\label{tab:bootstrap_rq1_jvr}
\begingroup
\fontsize{9}{10.5}\selectfont
\setlength{\tabcolsep}{3pt}
\renewcommand{\arraystretch}{1.0}
\newcommand{\bootstrapcell}[3]{%
  \begin{tabular}[c]{@{}c@{}}#1\\[-1pt]{\fontsize{8}{9.5}\selectfont [#2, #3]}\end{tabular}%
}

\begin{tabular*}{\linewidth}{@{}l@{\extracolsep{\fill}}ccccc@{}}
\toprule
Model
& \shortstack{$\Delta\mathrm{JVR}$\\{\fontsize{8}{9.5}\selectfont [95\% CI]}}
& \shortstack{Discordant\\tasks}
& $p_{\mathrm{Holm}}$
& \shortstack{First\\(\%)}
& \shortstack{Rank\\{\fontsize{8}{9.5}\selectfont [95\% interval]}} \\
\midrule
GPT-5.6-sol
& \bootstrapcell{+0.00}{+0.00}{+0.00}
& -- & -- & 84.5
& \bootstrapcell{1.0}{1.0}{2.5} \\
Claude-Fable-5.1
& \bootstrapcell{-1.29}{-3.72}{+0.97}
& 32 / 24 & 0.350 & 13.3
& \bootstrapcell{2.0}{1.0}{4.0} \\
Claude-Opus-5
& \bootstrapcell{-2.10}{-4.36}{+0.16}
& 35 / 22 & 0.222 & 2.0
& \bootstrapcell{3.0}{2.0}{4.0} \\
Kimi-K3
& \bootstrapcell{-3.23}{-5.65}{-0.81}
& 41 / 21 & 0.045 & 0.2
& \bootstrapcell{4.0}{2.0}{6.0} \\
Qwen3.8-max
& \bootstrapcell{-4.04}{-6.46}{-1.62}
& 42 / 17 & 0.006 & 0.0
& \bootstrapcell{5.0}{4.0}{7.0} \\
DeepSeek-V4-Pro
& \bootstrapcell{-4.68}{-7.11}{-2.26}
& 44 / 15 & 0.001 & 0.0
& \bootstrapcell{6.0}{4.0}{8.0} \\
DeepSeek-V4-Flash
& \bootstrapcell{-5.33}{-7.59}{-3.07}
& 44 / 11 & $<.001$ & 0.0
& \bootstrapcell{7.0}{5.5}{8.0} \\
GLM-5.3
& \bootstrapcell{-5.49}{-7.75}{-3.23}
& 45 / 11 & $<.001$ & 0.0
& \bootstrapcell{8.0}{5.5}{8.0} \\
\bottomrule
\end{tabular*}
\endgroup
\end{table}

\endgroup

\FloatBarrier

\subsection{Performance by Task Source}
Table~\ref{tab:split_leaders} expands the source comparison in RQ1 with the configuration and component rates for each source-specific JVR leader.
Each percentage is computed within the corresponding source split.
Issue composition and candidate-pool coverage vary across splits, with DeepSWE showing the lowest best JVR.
These differences preclude attributing the score gap to trajectory length from the split comparison alone.

\begin{table}[htbp]
\centering
\caption{
Configurations with the highest JVR in each task source.
}
\label{tab:split_leaders}
\vspace{0.4em}
\resizebox{\columnwidth}{!}{%
\begin{tabular}{llcccccc}
 \toprule
Source & Configuration & JVR & GRR & WDR & CAR & CT-WDR & CT-CAR \\
 \midrule
Multilingual ($n{=}167$) & GPT-5.6-sol & 16.17 & 87.43 & 82.57 & 79.19 & 38.92 & 50.30 \\
Pro ($n{=}342$) & Claude-Fable-5 & 8.77 & 67.25 & 84.23 & 66.19 & 30.70 & 32.16 \\
DeepSWE ($n{=}110$) & GPT-5.6-sol & 2.73 & 38.18 & 96.82 & 34.26 & 44.55 & 12.73 \\
 \bottomrule
 \end{tabular}
}%
 \end{table}

\FloatBarrier

\subsection{Qualitative Audits}
\label{app:audits}

\parabf{From a disagreement to its cause.}
Complementary studies diagnose failure development in coding-agent trajectories~\citep{li2026codetracertraceableagentstates,zhao2026failureprocessanatomycli}.
A disagreement audit collects the issue, repository, required behavior, candidate changes, generated assertions, and execution traces.
When a suite accepts an incorrect patch, we inspect which required behavior remains unchecked.
When it rejects a correct patch, we check whether the assertion reveals an oracle gap or unnecessarily constrains the implementation.
The collected pairs below establish patch diversity; their possible test failures illustrate audit questions and are not measurements of generated-suite errors.

\parabf{Repairing different layers: Redis.}
For \texttt{redis\_\_redis-13115}, Lua can format a numeric argument in scientific notation, but \texttt{HINCRBY} expects a decimal integer string.
Two correct patches have normalized cosine similarity 0.017: one changes formatting in \texttt{src/script\_lua.c}, while the other changes a conversion condition in \texttt{deps/fpconv/fpconv\_dtoa.c}.
Invoking Lua and checking the stored value would accommodate either repair location.
An assertion requiring a particular helper call would instead make acceptance depend on implementation structure.

\parabf{Repairing one function in different ways: Hugo.}
For \texttt{gohugoio\_\_hugo-12343}, URL parsing can treat the portion after \texttt{\#} as a fragment and truncate the path.
Two correct \texttt{PathEscape} patches have normalized cosine similarity 0.043: one escapes \texttt{\#} before parsing, while the other escapes individual path segments before reconstruction.
The shared edit location therefore does not imply a shared repair strategy.
The harness supplies both labels without proving equivalence on every possible path input.


\FloatBarrier

\section{SFT Data and Downstream Evaluation}
\label{app:sft}

SFT follows the model, teacher, and five-epoch training setup described in RQ2; downstream evaluation uses OpenHands.
Table~\ref{tab:sft_results} compares four training configurations: SWE SFT, + TestGen Supervision, + Validation-guided TestGen, and + Scaled TestGen Supervision.

\subsection{RQ2: Training Data and Selection}

\parabf{Data separation.}
We enforce repository-level disjointness between all SFT training data and evaluation tasks in \benchmark, SWE-bench Verified, SWE-bench Multilingual, and SWE-bench Pro.

\parabf{Training-data composition.}
The SWE SFT baseline contains 1.9 billion tokens of repository-level repair trajectories.
TestGen Supervision and Validation-guided TestGen each add 2 billion tokens of TestGen trajectories selected from the same pool, constructed from approximately 20,000 issue environments.
TestGen Supervision selects trajectories using GRR, whereas Validation-guided TestGen uses WDR/CAR/JVR-based selection.
Scaled TestGen Supervision adds a further 2 billion tokens of validation-selected TestGen trajectories from an expanded pool, bringing the total TestGen data to 4 billion tokens.

\parabf{Evaluation and selection protocol.}
The TestGen metrics in Table~\ref{tab:sft_results} are evaluated on the final 619-task benchmark.
Data selection follows the GRR and WDR/CAR/JVR criteria described in RQ2.

\parabf{What the training comparison isolates.}
The unscaled comparison holds the source trajectory pool and the 2-billion-token TestGen budget fixed while changing selection criteria.
Scaled supervision additionally changes the data quantity and environment pool, so its gains do not isolate selection from scaling or changes in data composition.

\subsection{RQ2: Test-Generation Reliability}
\label{app:statistical_reliability_rq2}

The RQ2 analysis covers the four SFT configurations reported in
Table~\ref{tab:sft_results}.
We apply the same paired, source-stratified bootstrap to their task-level
test-generation outcomes.
Table~\ref{tab:bootstrap_rq2_metrics} reports uncertainty intervals for
all eight \benchmark metrics.
Cells use the same percentage convention as Table~\ref{tab:bootstrap_rq1_metrics}, with point estimates matching Table~\ref{tab:sft_results} and intervals computed from unrounded task-level values.

\begin{table}[H]
\centering
\caption{
Bootstrap estimates and 95\% confidence intervals for the four RQ2 SFT configurations.
}
\label{tab:bootstrap_rq2_metrics}
\begingroup
\fontsize{9}{10.5}\selectfont
\setlength{\tabcolsep}{3pt}
\renewcommand{\arraystretch}{1.0}
\newcommand{\bootstrapcell}[3]{%
  \begin{tabular}[c]{@{}c@{}}#1\\[-1pt]{\fontsize{8}{9.5}\selectfont [#2, #3]}\end{tabular}%
}

\begin{tabular*}{\linewidth}{@{}l@{\extracolsep{\fill}}cccc@{}}
\toprule
\multicolumn{5}{@{}l}{\textbf{Reference and patch-level metrics}} \\[3pt]
Configuration & GRR & WDR & CAR & F1 \\
\midrule
SWE SFT
& \bootstrapcell{8.89}{6.79}{11.15}
& \bootstrapcell{20.75}{17.91}{23.66}
& \bootstrapcell{25.28}{22.10}{28.41}
& \bootstrapcell{22.79}{20.44}{25.08} \\
TestGen Supervision
& \bootstrapcell{39.26}{35.70}{42.81}
& \bootstrapcell{59.00}{56.40}{61.61}
& \bootstrapcell{43.98}{40.76}{47.16}
& \bootstrapcell{50.39}{48.05}{52.70} \\
Validation-guided TestGen
& \bootstrapcell{40.39}{36.83}{43.94}
& \bootstrapcell{64.37}{61.40}{67.27}
& \bootstrapcell{50.27}{47.03}{53.54}
& \bootstrapcell{56.45}{53.99}{58.81} \\
Scaled TestGen Supervision
& \bootstrapcell{41.20}{37.64}{44.91}
& \bootstrapcell{72.14}{69.62}{74.69}
& \bootstrapcell{51.62}{48.19}{55.05}
& \bootstrapcell{60.18}{57.75}{62.50} \\
\midrule
\addlinespace[3pt]
\multicolumn{5}{@{}l}{\textbf{Complete-task metrics}} \\[3pt]
Configuration & CT-WDR & CT-CAR & CT-F1 & JVR \\
\midrule
SWE SFT
& \bootstrapcell{9.21}{6.95}{11.47}
& \bootstrapcell{19.39}{16.48}{22.46}
& \bootstrapcell{12.49}{10.26}{14.56}
& \bootstrapcell{1.29}{0.48}{2.26} \\
TestGen Supervision
& \bootstrapcell{13.73}{11.15}{16.32}
& \bootstrapcell{28.11}{25.04}{31.18}
& \bootstrapcell{18.45}{15.86}{20.79}
& \bootstrapcell{2.26}{1.29}{3.39} \\
Validation-guided TestGen
& \bootstrapcell{25.69}{22.46}{29.08}
& \bootstrapcell{31.50}{28.27}{34.73}
& \bootstrapcell{28.30}{25.87}{30.65}
& \bootstrapcell{3.23}{1.94}{4.68} \\
Scaled TestGen Supervision
& \bootstrapcell{31.34}{27.79}{34.89}
& \bootstrapcell{32.96}{29.56}{36.35}
& \bootstrapcell{32.13}{29.77}{34.37}
& \bootstrapcell{3.88}{2.42}{5.49} \\
\bottomrule
\end{tabular*}
\endgroup
\end{table}

\parabf{Paired comparisons with the SFT baseline.}
We compare each test-generation training configuration with SWE SFT.
Here, $\Delta\mathrm{JVR}$ denotes the trained configuration's JVR minus
the SWE SFT baseline in percentage points under the paired stratified bootstrap.
We apply exact two-sided McNemar tests to the paired task-level JVR
outcomes and adjust the three planned comparison $p$-values using
Holm's procedure.
Table~\ref{tab:bootstrap_rq2_jvr} lists discordant task counts as baseline-only and compared-configuration-only JVR successes, respectively.
Rank stability is computed jointly over the four SFT configurations.

\begin{table}[H]
\centering
\caption{
Paired JVR comparisons and bootstrap rank stability for RQ2.
}
\label{tab:bootstrap_rq2_jvr}
\begingroup
\fontsize{9}{10.5}\selectfont
\setlength{\tabcolsep}{3pt}
\renewcommand{\arraystretch}{1.0}
\newcommand{\bootstrapcell}[3]{%
  \begin{tabular}[c]{@{}c@{}}#1\\[-1pt]{\fontsize{8}{9.5}\selectfont [#2, #3]}\end{tabular}%
}

\begin{tabular*}{\linewidth}{@{}l@{\extracolsep{\fill}}ccccc@{}}
\toprule
Configuration
& \shortstack{$\Delta\mathrm{JVR}$\\{\fontsize{8}{9.5}\selectfont [95\% CI]}}
& \shortstack{Discordant\\tasks}
& $p_{\mathrm{Holm}}$
& \shortstack{First\\(\%)}
& \shortstack{Rank\\{\fontsize{8}{9.5}\selectfont [95\% interval]}} \\
\midrule
SWE SFT
& \bootstrapcell{+0.00}{+0.00}{+0.00}
& -- & -- & 0.0
& \bootstrapcell{4.0}{3.0}{4.0} \\
TestGen Supervision
& \bootstrapcell{+0.97}{-0.16}{+2.10}
& 3 / 9 & 0.146 & 0.4
& \bootstrapcell{3.0}{2.0}{4.0} \\
Validation-guided TestGen
& \bootstrapcell{+1.94}{+0.81}{+3.23}
& 2 / 14 & 0.008 & 17.2
& \bootstrapcell{2.0}{1.0}{3.0} \\
Scaled TestGen Supervision
& \bootstrapcell{+2.58}{+1.29}{+4.04}
& 2 / 18 & 0.001 & 82.4
& \bootstrapcell{1.0}{1.0}{2.0} \\
\bottomrule
\end{tabular*}
\endgroup
\end{table}

The SFT configurations show consistent improvements over the SWE SFT
baseline across the test-generation metrics in
Table~\ref{tab:bootstrap_rq2_metrics}.
Scaled TestGen Supervision achieves the highest observed JVR at 3.88\%
and receives 82.4\% of the bootstrap first-place probability.
Its paired JVR improvement over SWE SFT is 2.58 percentage points, with
a 95\% confidence interval of $[1.29,4.04]$.
Validation-guided TestGen also improves JVR over the baseline, with a
paired difference of 1.94 percentage points and a 95\% confidence
interval of $[0.81,3.23]$.
These paired results support the gains of validation-guided and scaled supervision over SWE SFT.

\FloatBarrier

\subsection{RQ3: Downstream Evaluation}
AVG@4 and Pass@4 follow the four-round definitions in RQ3: mean per-round resolution and resolution in at least one round, respectively.
Table~\ref{tab:sft_results} reports AVG@4 and Pass@4 on SWE-bench Verified, Multilingual, and Pro.

\FloatBarrier

\section{Records Needed for Reproduction}
\label{app:reporting}

Reproducing a \benchmark comparison requires three groups of records:
\textbf{(1) Tasks and candidates:} identifiers, filtering decisions, source counts, producers, labels, final pool sizes, and empty-pool counts.
\textbf{(2) Executions and scores:} suite counts, commands, statuses, timeouts, replays, metric numerators and denominators, and per-task results for macro averaging.
\textbf{(3) Training and transfer:} checkpoint identities, data recipes, selection thresholds, retained trajectory manifests, and downstream attempt coverage.
Improvements in test generation or repair must be supported by execution outcomes, rather than training loss or judge preferences alone.

\section{Limitations}
\label{app:limitations}


\parabf{What the oracle and execution can establish.}
Candidate labels follow the official oracle, providing an operational notion of correctness rather than a formal guarantee of semantic correctness.
We mitigate oracle-quality concerns through task-level filtering of weak or implementation-dependent reference tests and manual quality checks.
Generated tests may nevertheless exercise behavior beyond the official oracle.
Timeouts, missing outputs, and applicability failures reduce fixed-denominator scores without necessarily evaluating behavioral assertions.
Execution records therefore help distinguish behavioral misclassification from unresolved execution.
Audits should connect test assertions and execution traces to issue requirements, distinguishing weaknesses in generated tests from gaps in official coverage.

\parabf{Which issues and patches are represented.}
The reproduction panel defines task difficulty, and its model families partly overlap those evaluated on the final set.
The selected challenge set therefore does not represent the full population of software issues.
The 82,579 candidate patches cover observed attempts, not every possible implementation or fault, and textual deduplication does not guarantee semantic diversity.
DeepSWE's thinner correct pools further constrain conclusions about correct-patch acceptance.

\end{document}